\documentclass[english,prb,twocolumn]{revtex4}
\usepackage{graphicx}
\usepackage{enumerate}
\usepackage{hyperref}
\usepackage{textcomp}
\usepackage{amsmath}
\usepackage{amssymb}
\usepackage{pgf}
\usepackage{subfigure}
\usepackage[english]{babel}
\usepackage{import}

\def\ud{\mathrm{d}}

\newcommand{\nep}{\textrm{e}}

\newcommand{\on}{0}

\renewcommand{\k}{{\bf k}}

\newcommand{\x}{{\bf x}}

\newcommand{\y}{{\bf y}}

\newcommand{\A}{{\bf A}}

\newcommand{\calA}{{\mathcal{A}}}

\newcommand{\currJ}{\widehat{\mathrm{J}}_y}
\newcommand{\currj}{\hat{\mathrm{j}}_y}
\newcommand{\opc}[1]{{\hat{c}^{\phantom \dagger}}_{#1}}
\newcommand{\opcdag}[1]{{\hat{c}^{\dagger}}_{#1}}

\begin{document}
\title{On the quantization of the Hall conductivity in the Harper-Hofstadter model}
\author{Matteo M. Wauters$^{1}$, Giuseppe E. Santoro$^{1,2,3}$}

\affiliation{
$^1$ SISSA, Via Bonomea 265, I-34136 Trieste, Italy\\
$^2$ International Centre for Theoretical Physics (ICTP), P.O. Box 586, I-34014 Trieste, Italy\\
$^3$ CNR-IOM Democritos National Simulation Center, Via Bonomea 265, I-34136 Trieste, Italy
}

\begin{abstract}
We study the robustness of the quantization of the Hall conductivity in the Harper-Hofstadter model towards 
the details of the protocol with which a longitudinal uniform driving force $F_x(t)$ is turned on. 
In the vector potential gauge, through Peierls substitution, this involves the switching-on of complex time-dependent hopping amplitudes
$\nep^{-\frac{i}{\hbar}\calA_x(t)}$ in the $\hat{\x}$-direction such that $\partial_t \calA_x(t)=F_x(t)$. 
The switching-on can be sudden, $F_x(t)=\theta(t) F$, where $F$ is the steady driving force,
or more generally smooth $F_x(t)=f(t/t_{\on}) F$, where $f(t/t_{\on})$ is such that $f(0)=0$ and $f(1)=1$. 
We investigate how the time-averaged (steady-state) particle current density $j_y$ in the $\hat{\y}$-direction 
deviates from the quantized value $j_y \, h/F = n$ due to the finite value of $F$ and the details of the switching-on protocol. 
Exploiting the time-periodicity of the Hamiltonian $\hat{H}(t)$, we use Floquet techniques to study this problem.
In this picture the (Kubo) linear response $F\to 0$ regime corresponds to the adiabatic limit for $\hat{H}(t)$. 
In the case of a sudden quench $j_y \, h/F$ shows $F^2$ corrections to the perfectly quantized limit.
When the switching-on is smooth, the result depends on the switch-on time $t_{\on}$: for a fixed $t_{\on}$ we observe a crossover
force $F^*$ between a quadratic regime for $F<F^*$ and a {\em non-analytic} exponential $\nep^{-\gamma/|F|}$ for $F>F^*$.
The crossover $F^*$ decreases as $t_{\on}$ increases, eventually recovering the topological robustness.
These effects are in principle amenable to experimental tests in optical lattice cold atomic systems with synthetic gauge fields. 
\end{abstract}

\maketitle

\section{Introduction}\label{sec:intro}
The quantization of the transverse conductivity $\sigma_{H}$ in the Integer Quantum Hall Effect~\cite{Klitzing_review} (IQHE) is probably the most famous 
manifestation of a topological invariant, the first Chern number, in condensed matter physics \cite{Niu_RMP10}. 
Indeed the celebrated TKNN paper~\cite{thouless1982quantized} showed that in the linear response regime, i.e. when the external electric field is small, 
the Hall conductivity predicted by the Kubo formula $j_y^{e} =\sigma_H E_x$ is quantized and can be written as
%
%
the sum of the Chern numbers of the occupied bands and therefore it must be an integer, in units of $e^2/h$. 

The extreme precision of the quantized Hall conductance revealed in the experiments~\cite{Klitzing_review} suggests a remarkable 
robustness of the IQH phase against many ingredients, notably the presence of impurities and interactions, and the strength of the applied electric field.
Concerning the latter issue, the mathematical physics literature~\cite{Klein_CIMP90} has shown that corrections to the Kubo formula vanish in 
Quantum Hall systems to all orders in perturbation theory.  

Quite recently, the issue of the topological robustness of a related phenomenon 
--- Thouless pumping in one-dimensional insulators \cite{Thouless_PRB83} --- has been re-examined, showing that the details of the preparation of the 
quantum non-equilibrium steady-state and of the time-interval in which the pumped charge is measured deeply influence how the topological 
$\tau\to \infty$ adiabatic limit is approached~\cite{Privitera_PRL18}. 
In particular, it was shown that the charge pumped over a {\em finite} number of periods shows non-analytic corrections
--- in the form of faster and faster oscillations as $\omega=2\pi/\tau\to 0$ --- when the periodic driving protocol is turned-on abruptly starting 
from an initial uncorrelated insulating state~\cite{Privitera_PRL18}.
Such a non-analytic approach of the adiabatic (topological) limit $\omega\to 0$ was indeed predicted by Avron \& Kons~\cite{Avron_JPA99} through 
rigorous general arguments.  
What such rigorous arguments do not tell is how the limit $\omega\to 0$ is approached when one considers the asymptotic (steady state) 
single-period pumped charge, where topological effects should most appropriately looked for \cite{Avron_PT03}, because this involves an {\em infinite-time limit}. 
Remarkably, Ref.~\onlinecite{Privitera_PRL18} shows that non-analytic corrections present at finite-time turn into {\em quadratic} corrections $\sim\omega^2$
when the asymptotic pumped charge is considered. 
 
Modern realizations of the IQHE physics involve artificial gauge fields in cold atomic systems~\cite{Bloch_RevModPhys08,Aidels_PRL13,Miyake_PRL13,Bloch_Nphys15}.
In the light of the results of Ref.~\onlinecite{Privitera_PRL18}, these experiments raise the non-trivial issue of the robustness of the quantized Hall conductance 
against many details, including primarily the preparation of the Quantum Hall state and the ensuing turning-on of the constant field, as well as the measurement 
of the transverse current. 
To set up and state the problem we will address, let us assume that the coherence-time \cite{Bloch_RevModPhys08} of these cold atomic systems is
so long that it is legitimate to estimate the time-average transverse current from its infinite-time limit  
\begin{equation} \label{eqn:jy_intro}
j_y=\lim_{T\to\infty} \frac{1}{T}\int_{t_{\on}}^{t_{\on}+T} \! \ud t' \, \langle \psi(t') | \, \hat{\rm j}_y | \psi(t') \rangle \;.
\end{equation}
Here $\hat{\rm j}_y$ is the space-averaged particle-current density operator, and $|\psi(t)\rangle$ is assumed to evolve unitarily with the system Hamiltonian $\hat{H}(t)$, 
including the external uniform force field $F_x(t)$ in the $\hat{\x}$-direction, which we represent by an extra time-dependent vector potential 
$\calA_x \hat{\x}$ with $\partial_t \calA_x(t)=F_x(t)$.
Furthermore, let us assume that the external uniform force $F_x(t)$ is switched-on in a time $t_{\on}$ towards a stationary value $F$, i.e.,
$F_x(t)=f(t/t_{\on}) F$, where $f(t/t_{\on})$ is a switching-on function with $f(0)=0$ and $f(1)=1$.
The Kubo formula for the IQHE implies that for small $F$:
\begin{equation} \label{eq:LinResp}
j_y  = \sigma_{yx} F = \frac{n}{h} F \;,
\end{equation}
meaning that in the limit $F\rightarrow 0$ the quantity $j_y h/F$ is {\em exactly} an integer number $n$. 
A robust quantization against the strength of $F$ would appear, in this context, as non-analytic corrections of the form $j_y h/F \simeq n + A \nep^{-\gamma/|F|}$,
while the presence of quadratic corrections, $j_y h/F \simeq n + B F^2 + o(F^2)$, would signal an ordinary perturbative response. 

In this paper we investigate how the finite value of the stationary driving force $F$ and the details of the driving protocol, encoded in $t_{\on}$ and 
in the switching-on function $f(s=t/t_{\on})$, affect the precision of the measurement of the transverse Hall response in cold atoms IQHE systems. 
%
\begin{figure}[h]
\begin{center}
\def\svgwidth{75mm}
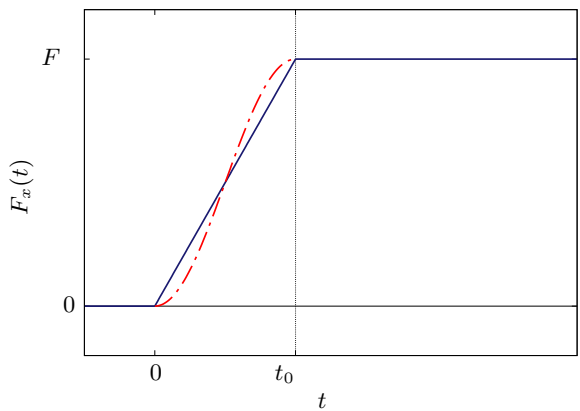
\caption{Possible schedules for the switching-on of the uniform driving force $F_x(t)$.
The sudden quench case $F_x(t)=\theta(t) F$, where $\theta(t)$ is the Heaviside step function, is recovered for $t_{\on} \to 0$.
}
\label{fig:Fx}
\end{center}
\end{figure}
Our investigation focuses on the Harper-Hofstadter (HH) model~\cite{Hofstadter_PRB76}, 
a two-dimensional tight-binding Hamiltonian for IQHE which is particularly relevant for experimental realizations with optical lattices~\cite{Aidels_PRL13,Miyake_PRL13,Bloch_Nphys15}, providing an excellent tool to study QHE physics in a tunable and controlled system.
The techniques used involve quite standard Floquet tools to study the time-periodic dynamics of the transverse current, which can be formulated as 
a quantum pumping problem. 

We will show that the main responsible for topological robustness is the switching function $f(t/t_{\on})$.
We analyze in detail three possible schedules: a sudden quench $F_x(t)=\theta(t) F$, a linear ramp $F_x(t)=(t/t_{\on}) F$ and a smoother ramp,
as sketched in Fig.~\ref{fig:Fx}.
We will show that in the sudden case, $F_x(t)=\theta(t) F$, the Hall response of the system is perturbative, and $F^2$ corrections to $j_y h/F$ are present.  
When the driving force is turned-on linearly in a time $t_{\on}$, $F_x(t)=(t/t_{\on}) F$, we find two distinct regimes: 
for a fixed $t_{\on}$ we observe a crossover force $F^*(t_{\on})$ between a quadratic regime for $F<F^*$ and a {\em non-analytic} exponential 
$\nep^{-\gamma/|F|}$ for $F>F^*$. The crossover $F^*(t_{\on})$ decreases as $t_{\on}$ increases, eventually recovering the topological robustness.
Finally, if the switching-on is smoother (with a continuous derivative), $F_x(t)=\frac{1}{2} \left(1-\cos(\pi t/t_{\on}) \right) F$, 
we observe no qualitative differences with the linear ramp case, suggesting the main ingredient for the topological robustness seems to be the continuity of $F_x(t)$
and a suitably long $t_{\on}$. 
%

The paper is organized as follows:
in Sec.~\ref{model:sec} we introduce the Harper-Hofstadter and the quantum pumping approach we used to study the non-adiabatic corrections to the transverse response.
In Sec.~\ref{sec:results} we present our results, along with a detailed analysis of the topological nature of the pumped charge, highlighting the main factors responsible for the correction to the quantized transverse response.
In particular, in Sec.~\ref{sseq:floq} we analyze the topological properties of the transverse current carried by a Floquet state,
in Sec.~\ref{sseq:quench} we describe the response to the sudden quench of the external force 
and in Sec.~\ref{sseq:ramp} we discuss the continuous switching-on of $F_x(t)$.
Conclusions and outlook are contained in Sec.~\ref{sec:conclusion}.

\section{The Harper-Hofstadter model} \label{model:sec}
%
Our starting point is the Harper-Hofstadter Hamiltonian \cite{Hofstadter_PRB76},  which describes a tight-binding system of 
non-interacting spinless fermions on a two-dimensional (2D) square lattice, pierced by a uniform magnetic field 
${\bf B}=B\hat{{\bf z}}$ perpendicular to the lattice plane:
\begin{equation} \label{eq:harper_H_0}
\hat{H} = - J_0 \sum_{l,m} \Big[ \opcdag{l+1,m} \opc{l,m} + \nep^{-i2\pi \alpha l} \opcdag{l,m+1} \opc{l,m} + {\mathrm H.c.} 
\Big] \;.
\end{equation}
Here $J_0$ is the bare hopping amplitude, and $(l,m)$ are integers labelling the square lattice sites, ${\bf r}_{l,m}=a(l\hat{\x} +m\hat{\y})$,
with lattice spacing $a$, with boundary conditions to be discussed later on. 
The magnetic field flux per plaquette, in units of the flux quantum $\phi_0=hc/e$, is here $\alpha=a^2B/\phi_0$, and
results in a complex hopping amplitude through Peierls' substitution, 
$J_{0} \, \nep^{-i \frac{e}{\hbar c} \int_{{\bf r}}^{{\bf r}'} {\bf A} \cdot \ud {\bf x}}$, 
with a Landau gauge choice for the vector potential $\A=Bx \hat{{\bf y}}$, breaking translational invariance along the 
$\hat{\x}$-direction.
In a condensed matter realization of this model Hamiltonian, with charged particles in real magnetic fields, 
one would not be able to explore the full phase diagram of the model for $\alpha\in [0,1]$, since the flux per plaquette 
is too small, even with large laboratory fields.  
In modern realizations with neutral cold atoms in optical lattices \cite{Dalibard_RMP11, Bloch_Nphys15}, on the contrary, 
synthetic gauge fields are used and all interesting values of $\alpha$ are possible. 
Historically, as discovered by Hofstadter \cite{Hofstadter_PRB76}, the spectrum is extremely complex, with rational values of $\alpha=p/q$ 
leading to $q$ energy sub-bands with gaps in between.
The crucial realization, due to Thouless and coworkers \cite{thouless1982quantized}, is that the insulating states
obtained when the Fermi energy lies inside the gaps between such sub-bands has a quantized Hall conductance
\begin{equation}\label{eq:TKNN}
\sigma_{H}=-\frac{e^2}{h}\sum_{\nu}^{\rm occ} \int_{\rm BZ} \frac{\ud^2 \mathbf{k}}{2\pi} \Omega_{\nu}(\mathbf{k}) = n \frac{e^2}{h} \;,
\end{equation}
where $\Omega_{\nu}(\mathbf{k}) = i\Big[ \langle \partial_{k_x} u_{\nu,\mathbf{k}}|\partial_{k_y} u_{\nu,\mathbf{k}}\rangle - 
\langle \partial_{k_y} u_{\nu,\mathbf{k}}|\partial_{k_x} u_{\nu,\mathbf{k}}\rangle \Big]$ is the Berry curvature \cite{Niu_RMP10} 
of $\nu$-th occupied band, and $u_{\nu,\mathbf{k}}$ denote the periodic part of the Bloch wave-functions on the (magnetic) 
Brillouin Zone (BZ) of the system. 
This implies that a Hall current flows, for instance, in the $y$-direction when an electric field $E_x$ acts in the $x$-direction: $j_y=\sigma_{H} E_x$.
The robustness of this phenomenon is remarkable: disorder and (weak) interactions do not alter the result, thus providing an exceptionally
precise measurement~\cite{Klitzing_review} of $e^2/h$. 
A further remarkable robustness is offered by the fact that the Kubo-formula, derived from linear response theory, seems to extend its 
regime of validity well beyond linear response: as mathematically proven in Ref.~\onlinecite{Klein_CIMP90}, and further discussed in 
Ref.~\onlinecite{Avron_JPA99}, all power-law corrections in the electric field can be shown, under suitable hypotheses, to be missing.

The availability of new experiments employing synthetic gauge fields \cite{Dalibard_RMP11, Bloch_Nphys15}, directly sensitive to the 
time-dependent transient leading to the transverse response, calls for a further scrutiny of this issue. 
Experimentally, the driving force $F_x(t)$ in the $\hat{\bf x}$-direction can be turned on, as a function of time, with some freedom,  
either abruptly or in a more or less smooth fashion. 
On the theory side, we can represent such a force in different gauges: quite conveniently, for a finite-length system $L_x$ 
with periodic boundary conditions (PBC) in the $\hat{\x}$-direction, 
we can choose a vector-potential gauge in which the force is represented by a time-dependent vector potential. 
The minimal coupling requires, in a tight-binding scheme, the Peierls' substitution:
\begin{equation} \label{eq:Peierls_x}
\opcdag{l+1,m} \opc{l,m} \longrightarrow \nep^{-ia \kappa_x(t)} \, \opcdag{l+1,m} \opc{l,m} \;,
\end{equation}
where $\kappa_x(t)$ 
determines the force $F_x(t)$ acting in the $x$-direction through $F_x(t)=\hbar\dot{\kappa}_x(t)$, 
hence making the Hamiltonian time-dependent, $\hat{H}(t)$.
More in detail, we chose $F_x(t) = F f(t/t_{\on})$, where $f(s=t/t_{\on})$ is a switch-on function interpolating between $0$ and $1$, 
i.e., such that $f(s\le 0)=0$ and $f(s\ge 1)=1$, and $F$ is the stationary value of the force, attained for $t\ge t_{\on}$.
This choice leads to $\kappa_x(t\le t_{\on})=(t_{\on} F/\hbar) \int_{\on}^{t/t_{\on}} \ud s f(s)$ and 
$\kappa_x(t\ge t_{\on})=\kappa_x(t_{\on}) + F (t-t_{\on})/\hbar$. 
The case of a sudden switch-on of the force is recovered by taking $t_{\on}=0$.  
Since $\kappa_x(t)$ appears in the hopping as a phase-factor, see Eq.~\eqref{eq:Peierls_x}, its linear increase for $t\ge t_{\on}$ implies 
that the Hamiltonian becomes {\em time-periodic} for $t\ge t_{\on}$, $\hat{H}(t+\tau)=\hat{H}(t)$, with the period $\tau$ given by
\begin{equation} \label{eqn:tau_F_relation}
\tau = \frac{2\pi \hbar}{aF} \;,
\end{equation} 
which corresponds to a fundamental frequency $\hbar\omega=aF$ entering the problem. 
These considerations clearly show that the question of the validity of linear response in $F$ goes hand-in-hand with the issue of
{\em adiabaticity} of $\hat{H}(t)$: Kubo linear response is essentially obtained in the fully adiabatic limit $\omega\to 0$. 

%
To calculate the current, following Laughlin \cite{Laughlin_PRB81}, we use PBC in the $\hat{\bf y}$-direction as well, 
introducing a vector potential, again with a minimal-coupling Peierls' substitution:
\begin{equation} \label{eq:Peierls_y}
\opcdag{l,m+1} \opc{l,m} \longrightarrow \nep^{-ia \kappa_y} \, \opcdag{l,m+1} \opc{l,m} \;.
\end{equation}
The total current operator is obtained as a derivative of $\hat{H}$ with respect to $\kappa_y$:
\begin{equation}
\currJ = \frac{1}{\hbar} \frac{\partial \hat{H}}{\partial \kappa_y} \bigg|_{\kappa_y=0} \;.
\end{equation}
The Hall response can now be seen as a non-vanishing quantum average of $\currJ$ in presence of a force $F_x$, 
describing the transport of particles along $\hat{\bf y}$-direction.  
We can quantify this through the linear-density of transported particles during the interval $[t_{\on},t]$ 
(dropping the initial  switching-on interval $[0,t_{\on}]$):
\begin{equation} \label{eq:Q_y} 
Q_y(t\ge t_{\on}) =  \int_{t_{\on}}^{t} \ud t'  \; \langle \psi(t') |\, \currj |\psi(t')\rangle \;,
\end{equation}
where $|\psi(t)\rangle$ denotes the time-evolving state of the system, and $\currj=\currJ/(L_xL_y)$ is the space-averaged current density.
%
Notice that $Q_y$, as defined, gives the number of particles per-unit-length moving along the $\hat{\bf y}$-direction in the interval $[t_{\on},t]$:
we will often refer to it as {\em pumped charge}, although the particles could be neutral.   

We can ask for the charge pumped in the $m$-th period: 
\begin{equation} \label{eq:Q_m}
Q_m = Q_y(t_{\on}+m\tau) - Q_y(t_{\on}+(m-1)\tau) \;.
\end{equation}
We expect that the charge pumped in the initial periods $Q_1, Q_2, \cdots$ might be affected by {\em transient effects}, depending
on the details of the switching-on function $f(t/t_{\on})$ and time $t_{\on}$. 
These transient effects are expected to decay for $m\to \infty$, so that the infinite-time average
\begin{equation} \label{eqn:Q_average}
\overline{Q} = \lim_{M\to \infty} \frac{1}{M} \sum_{m=1}^M Q_m \;,
\end{equation}
should effectively capture the asymptotic (steady state) single-period pumped charge, where topological effects should  
most appropriately looked for \cite{Avron_PT03}.
The Floquet theorem enormously simplifies the calculation of the infinite-time average $\overline{Q}$. 
Indeed, the state of the system at any time $t\ge t_{\on}$ 
can be expanded in terms of Floquet modes and quasi-energies \cite{Grifoni1998driven,Holthaus_JPB16} as:
\begin{equation}
|\psi(t)\rangle = \sum_{\nu} \nep^{-i\epsilon_{\nu} (t-t_{\on})/\hbar} |u_{\nu}(t)\rangle \langle u_{\nu}(t_{\on}) |\psi(t_{\on})\rangle 
\end{equation}   
where $\epsilon_{\nu}$ are the Floquet quasi-energies and $|u_{\nu}(t)\rangle$ the associated time-periodic Floquet modes,  
$|u_{\nu}(t+\tau)\rangle=|u_{\nu}(t)\rangle$.
A rather standard derivation \cite{Avron_JPA99,Russomanno_PRL12} shows that the infinite-time average pumped charge is
dominated by the {\em Floquet diagonal ensemble} value:
\begin{equation} \label{eqn:Qbar_Qd} 
\overline{Q} \equiv Q_{\rm d} = 
\sum_{\nu} n_{\nu} \int_{t_{\on}}^{t_{\on}+\tau} \hspace{-5mm} \ud t'  \, \langle u_{\nu}(t') | \, \currj | u_{\nu}(t') \rangle \;.
\end{equation}
where $n_{\nu}=| \langle u_{\nu}(t_{\on}) | \psi(t_{\on})\rangle |^2$ is the ``occupation'' of the $\nu$-th Floquet mode. 
This clearly shows that the initial preparation, with the transient loading interval $[0,t_{\on}]$, is all contained in the occupation factors $n_{\nu}$. 

So far, we have made use of time-periodicity, but not of translational invariance. 
To proceed, we make a rational choice of the magnetic flux, $\alpha=p/q$ with $p$ and $q$ co-prime integers, 
which leads to an enlarged ``magnetic'' unit cell of size $qa$ in the $x$-direction. 
We now label the sites in the $x$-direction with a cell-index $j=0\cdots N_x-1$ and an intra-cell index $b=0,1,\cdots q-1$, so that $l=qj+b$,
while $m=0\cdots N_y-1$ labels sites in the $y$-direction. Hence, $L_x=N_x qa$, and $L_y=N_y a$.
We then define appropriate Bloch combinations of the form:
\begin{equation}
\left\{ 
\begin{array}{l}
\displaystyle \opcdag{\k,b} = \frac{1}{\sqrt{N}} \sum_{j=0}^{N_x-1} \sum_{m=0}^{N_y-1} \nep^{ia (k_x (qj+b) + k_y m)} \opcdag{qj+b,m}  \\
\displaystyle \opcdag{qj+b,m} = \frac{1}{\sqrt{N}} \sum_{\k}^{{\rm BZ}} \nep^{-ia (k_x (qj+b) + k_y m)} \opcdag{\k,b}
\end{array} \right. \;,
\end{equation}  
where $\k=\frac{2\pi}{a} \left( \frac{n_x}{qN_x} \hat{\x} + \frac{n_y}{N_y} \hat{\y} \right)$, with $n_x=0,\cdots,N_x-1$ and $n_y=0,\cdots,N_y-1$,
define the $N=N_xN_y$ wave-vectors inside the Brillouin Zone (BZ): $[0,\frac{2\pi}{qa}]\times [0,\frac{2\pi}{a}]$. 
The Hamiltonian for the system can then be written in the form:
\begin{eqnarray} \label{eq:harper_H_k}
\hat{H}(t) &=& - J_0 \sum_{\k}^{\rm BZ} \sum_{b=0}^{q-1} 
\bigg\{ 2 \cos\left( ak_y + {\scriptstyle{\frac{2\pi p}{q}}} b \right) \opcdag{\k,b} \opc{\k,b} \nonumber \\ 
&& \hspace{6mm}  + \Big[  \nep^{-ia(k_x+\kappa_x(t))} \opcdag{\k,b+1} \opc{\k,b} + {\mathrm H.c.}  \Big] \bigg\} \nonumber \\
 &=& \sum_{\k}^{\rm BZ} ( \opcdag{\k,0} \cdots \opcdag{\k,q-1} ) \cdot  {\mathbb H}(\k,t) \cdot  
 \left(  \begin{array}{l} \opc{\k,0} \\ \vdots \\ \opc{\k,q-1} \end{array} \right) \;, \hspace{4mm}
\end{eqnarray}
{\em i.e.}, effectively a $q\times q$ matrix problem ${\mathbb H}(\k,t)$ for every $\k$-vector in the BZ. 
The total current operator has a similar expression:
\begin{eqnarray} \label{eq:harper_J_k}
\currJ &=& \frac{2 a J_0}{\hbar} \sum_{\k}^{\rm BZ} \sum_{b=0}^{q-1} 
\sin\left( ak_y + {\scriptstyle{\frac{2\pi p}{q}}} b \right) \opcdag{\k,b} \opc{\k,b} \nonumber \\
&=& \sum_{\k}^{\rm BZ} ( \opcdag{\k,0} \cdots \opcdag{\k,q-1} )  \cdot {\mathbb J}(\k) \cdot  
 \left(  \begin{array}{l} \opc{\k,0} \\ \vdots \\ \opc{\k,q-1} \end{array} \right) \;,
\end{eqnarray}
where ${\mathbb J}(\k)=(1/\hbar) {\partial {\mathbb H}}/{\partial k_y}$. 

From now on, we will concentrate our study on the case $\alpha=1/3$, where the Hamiltonian becomes a $3\times 3$ problem for every $\k$.   
Exploiting the $\k$-factorization of the initial state $|\psi(0)\rangle$ and of the subsequent dynamics, 
using that the space-averaged current density is $\currj=\currJ/(L_xL_y)$,  
and transforming the sum over $\k$ into an integral on the BZ in the usual fashion for a large system,  
we can rewrite the asymptotic pumped charge as:
\begin{equation} \label{eq:Q_diag_orig}
Q_{\rm d} =  \sum_{\nu} \int_{\rm BZ} \frac{\ud^2 \k}{(2\pi)^2} n_{\k,\nu} 
\int_{t_{\on}}^{t_{\on}+\tau} \hspace{-5mm} \ud t'  \, \langle u_{\k,\nu}(t') | \frac{1}{\hbar} \frac{\partial {\mathbb H}}{\partial k_y} | u_{\k,\nu}(t') \rangle \;,
\end{equation}
where
\begin{equation}
n_{\k,\nu}=\big| \langle u_{\k,\nu}(t_{\on}) | \psi_{\k}(t_{\on})\rangle \big|^2 \;.
\end{equation}
A generalization of the Hellman-Feynman theorem for the Floquet case \cite{Grifoni1998driven}
shows that the average current carried by a Floquet mode is easily expressed in terms of the quasi-energy velocity:
\begin{equation}
\int_{t_{\on}}^{t_{\on}+\tau} \hspace{-5mm} \ud t'  \, \langle u_{\k,\nu}(t') | \frac{\partial {\mathbb H}}{\partial k_y} | u_{\k,\nu}(t') \rangle
= \tau \frac{\partial \epsilon_{\k,\nu}}{\partial k_y} \;.
\end{equation}
Hence $Q_{\rm d}$ in Eq.~\eqref{eq:Q_diag_orig} can be re-expressed as:
%
\begin{equation} \label{eq:Q_diag}
Q_{\rm d} = \frac{\tau}{\hbar} \sum_{\nu} \int_{\rm BZ} \frac{\ud^2 \k}{(2\pi)^2} \; 
n_{\k,\nu} \frac{\partial \epsilon_{\k,\nu}}{\partial k_y} \;.
\end{equation}
Henceforth we will refer to the quantity defined in Eq.~\eqref{eq:Q_diag} as the {\em diagonal pumped charge}.
We should stress that both the occupation factors $n_{\k,\nu}$ and the quasi-energies $\epsilon_{\k,\nu}$ are dependent on the 
time-periodicity $\tau$, or the frequency $\omega$, a dependence that we have not explicitly indicated. 

It is useful to make here the connection with the time-averaged current density mentioned in the introduction, Eq.~\eqref{eqn:jy_intro}. 
Using Eq.~\eqref{eqn:Q_average} and the fact that $\overline{Q} \equiv Q_{\rm d}$, see Eq.~\eqref{eqn:Qbar_Qd}, it is straightforward to derive that
\begin{equation} \label{eqn:jy_Q}
j_y=\lim_{T\to\infty} \frac{1}{T}\int_{t_{\on}}^{t_{\on}+T} \! \ud t' \, \langle \psi(t') | \, \hat{\rm j}_y | \psi(t') \rangle \equiv \frac{Q_{\rm d}}{\tau} \;.
\end{equation}
Hence, using the relationship \eqref{eqn:tau_F_relation} between the period $\tau$ and the force $F$, it is simple to show that:
\begin{equation}
j_y = \sigma_{yx} F = \frac{a Q_{\rm d}}{h} F  \;. 
\end{equation}
Hence, the transverse Hall ``conductance'' is here given by $\sigma_{yx} = (a Q_{\rm d})/h$. Its quantization, in units of $1/h$, would require
that $a Q_{\rm d} = n$, an integer. 

A final comment concerning transient effects. To appreciate them, the asymptotic pumped charge $Q_{\rm d}$ should be contrasted with the charge 
pumped in the $m$-th period, which would read:
\begin{equation}
Q_{m} = \frac{1}{\hbar} \int_{\rm BZ} \frac{\ud^2 \k}{(2\pi)^2} \; \int_{t_{\on}+(m-1)\tau}^{t_{\on}+m\tau} \hspace{-5mm} \ud t'
\, \langle \psi_{\k}(t') | \frac{\partial {\mathbb H}}{\partial k_y} | \psi_{\k}(t') \rangle \;.
\end{equation}
%

\section{Results} \label{sec:results}
\begin{figure}[ht]
\begin{center}
\def\svgwidth{75mm}
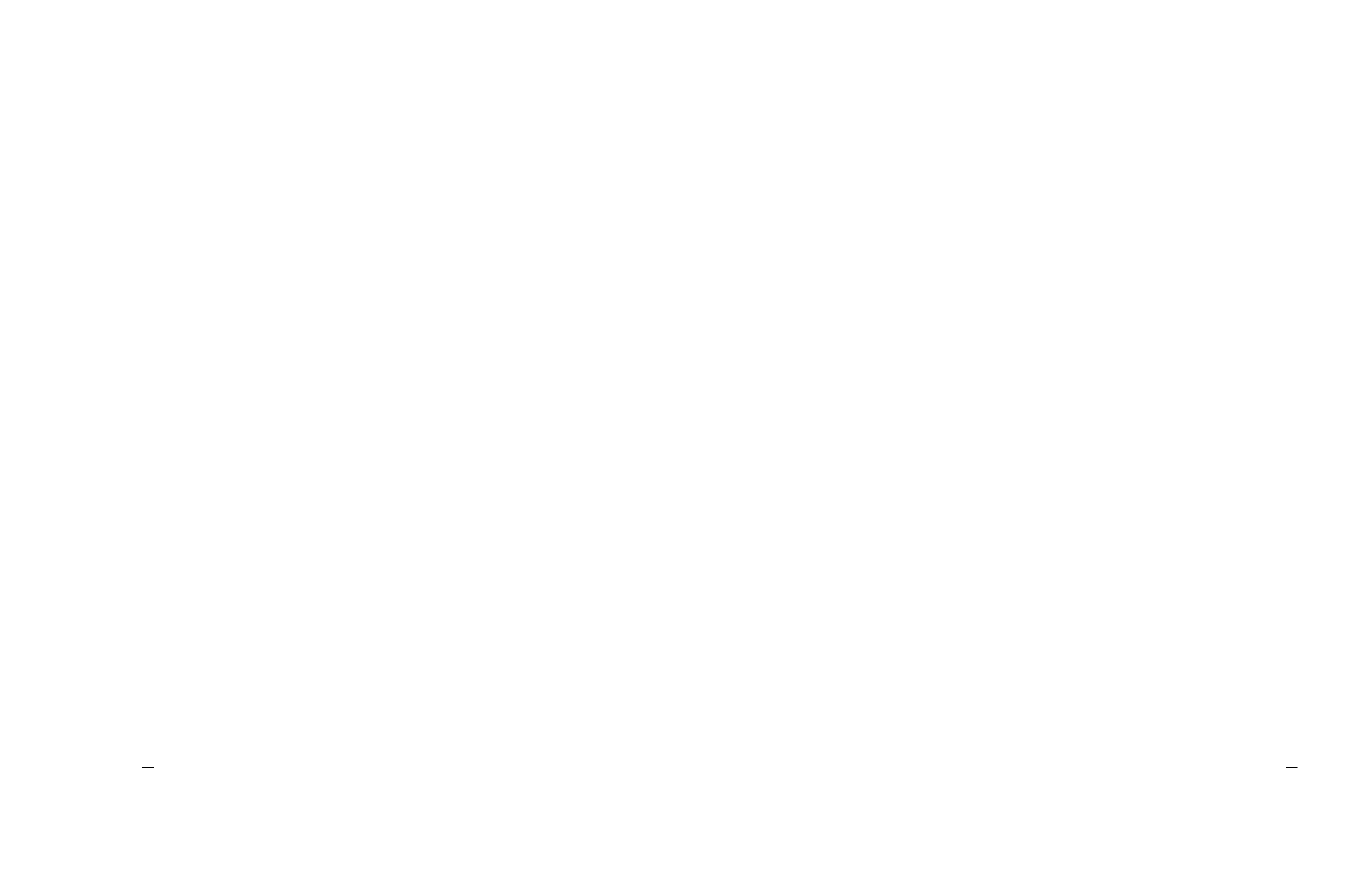\\
\def\svgwidth{75mm}
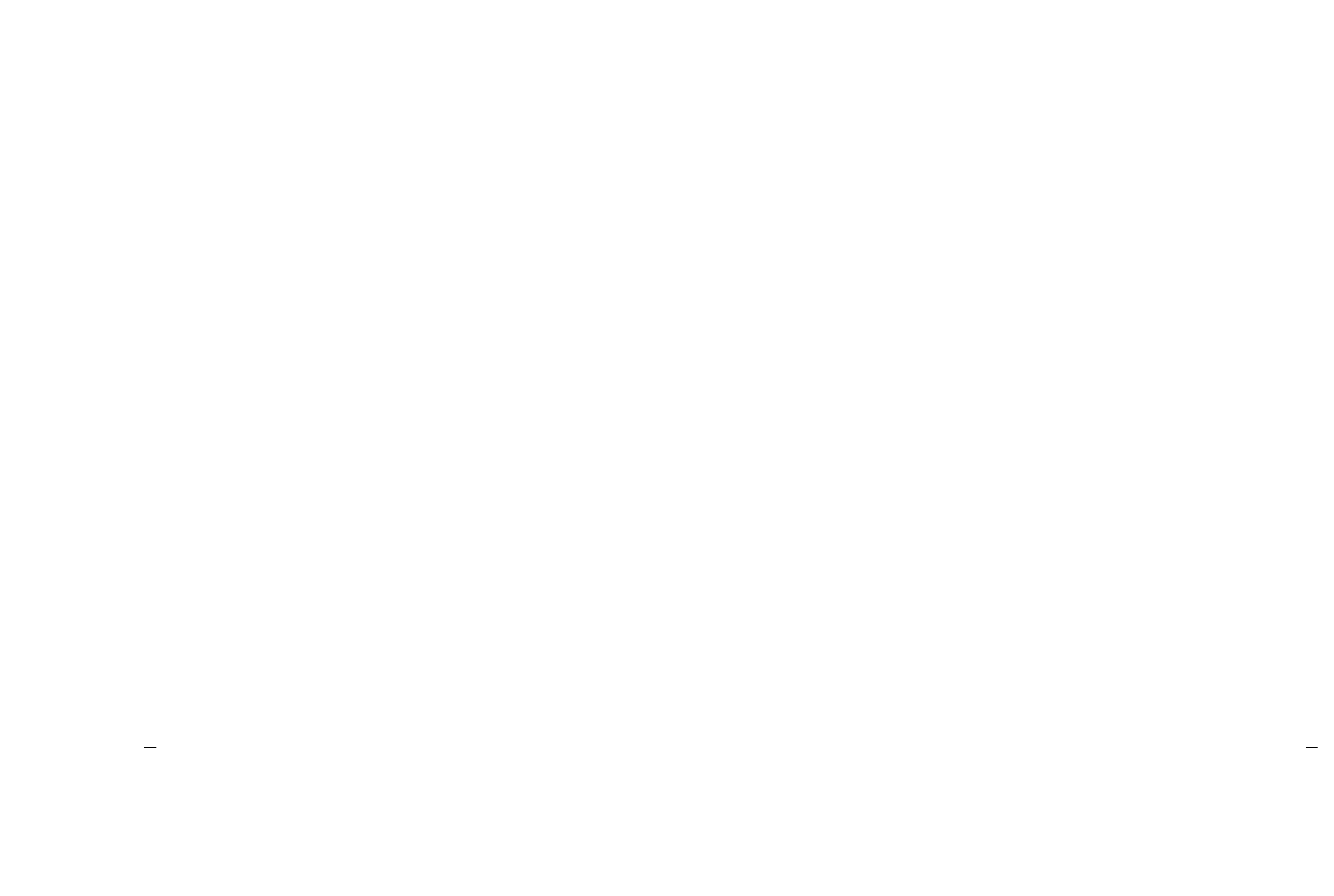
\caption{Top: Energy bands of the Harper-Hofstadter model for $\alpha=1/3$ vs $k_y$ for $N_x=30$.
The dashed red lines represent the energy averaged over the phase $k_x+\kappa_x(t)$.
Bottom: Charge pumped in the first period in each magnetic cell, $3aQ_1$, as a function of the driving field $aF/J_0$, where $\hbar\omega=aF$,
after a sudden switch-on of the driving. 
For $F\rightarrow 0$, $3a Q_1$ is quantized to the first Chern number, respectively +3, -6, +3, of the band in which the system is initially prepared. 
(The simulation has been repeated by preparing the initial Slater determinant insulating state to be one of the three completely filled bands, 
in order to compute the Chern numbers.)
The first and the third band give exactly the same response. 
This figure is essentially equivalent to Figure 1 of Ref.~\onlinecite{Avron_JPA99}, where the abscissa is $1/\omega$.
}
\label{fig:bands-chern}
\end{center}
\end{figure}

%
To illustrate the previous general considerations, let us consider the case of a sudden switch-on of a constant force $F_x(t)=\theta(t) F$,
which effectively amounts to taking $t_{\on}=0$ in the previous expressions. 
The bands of the unperturbed Hamiltonian are shown in Fig.~\ref{fig:bands-chern} (top), for $N_x=30$, as a function of $k_y\in (-\pi/a,\pi/a]$: 
we see $q=3$ distinct bands, obtained by projecting the $N_x$ different values of $k_x$. 
The initial insulating state is a Slater determinant $|\psi(0)\rangle$ obtained by completely filling one such band, for instance the lowest one.
We then calculate the charge pumped in the first period:
\begin{equation} \label{eq:Q_1}
Q_{1} = \frac{1}{\hbar} \int_{\rm BZ} \frac{\ud^2 \k}{(2\pi)^2} \; \int_{0}^{\tau} \hspace{-2mm} \ud t' \,
\langle \psi_{\k}(t') | \frac{\partial {\mathbb H}}{\partial k_y} | \psi_{\k}(t') \rangle \;.
\end{equation}
Fig.~\ref{fig:bands-chern} (bottom) shows the value of $Q_1$ as a function of the driving $F$, expressed in terms of  
$\hbar\omega=aF$. 
%
Notice that for $F\to 0$ we recover, as expected, a pumped charge which is quantized to the integer Chern numbers (+3,-6 and +3) 
of the three completely filled bands. Deviations from perfect quantization are clearly visible at finite $\omega$: the remaining part of
the paper is precisely devoted to understanding the nature and size of these deviations. 

In order to proceed with the analysis of the deviations from perfect quantization for small $\omega$, we shift our attention to the 
infinite-time average of the pumped charge, where the Floquet theory helps in elucidating the crucial ingredients. 
Our starting point is hence Eq.~\eqref{eq:Q_diag}, which we rewrite below for convenience in a slightly different form:
\begin{equation} \label{eq:Q_diag_bis}
Q_{\rm d} = \frac{1}{\hbar\omega} \sum_{\nu} \int_{\rm BZ} \frac{\ud^2 \k}{2\pi} \; 
n_{\k,\nu} \frac{\partial \epsilon_{k_y,\nu}}{\partial k_y} \;.
\end{equation}
In this re-writing we have used the trivial fact that $\tau=2\pi/\omega$ and that the Floquet quasi-energies $\epsilon_{\k,\nu}$
{\em dependent only on $k_y$}:  to appreciate the last fact, observe that the dependence of the Hamiltonian, see Eq.~\eqref{eq:harper_H_k},
on $k_x$ and $t$ is all contained in the phase-factor 
$\nep^{-ia(k_x+\kappa_x(t))}=\nep^{-ia\kappa_x(t_{\on})} \nep^{-i \omega (t-t_x)}$,
with $t_x=t_{\on}-ak_x/\omega$.
Hence, different values of $k_x$ effectively correspond to a shift in time $t_{\on}\to t_{\on}-t_x$, which in turn amounts to a unitary transformation on the
Floquet operator $\hat{U}_{\k}(t_{\on}+\tau,t_{\on})$, whose eigenvector/eigenvalues are the Floquet modes/quasi-energies:
\begin{equation} 
\hat{U}_{\k}(t_{\on}+\tau,t_{\on}) |u_{\k,\nu}(t_{\on}) \rangle = \nep^{-i\epsilon_{\k,\nu} \tau/\hbar} |u_{\k,\nu}(t_{\on})\rangle \;. 
\end{equation}
As discussed in App.~\ref{app:Floquet}, the fact that 
the Floquet operators at different $k_x$ are {\em unitarily equivalent} implies that their eigenvalues are $k_x$-independent, i.e.,
$\nep^{-i\epsilon_{k_y,\nu} \tau/\hbar}$. 
Notice that, on the contrary, the Floquet modes $|u_{\k,\nu}(t)\rangle$, and hence the occupations 
$n_{\k,\nu}=| \langle u_{\k,\nu}(t_{\on}) | \psi_{\k}(t_{\on})\rangle |^2$, {\em do depend} on $k_x$. 

\subsection{Pumping of Floquet states.}\label{sseq:floq}
The first issue we address is what happens to the pumped charge if the initial state $|\psi(t_{\on})\rangle$ is 
{\em precisely prepared to be the $\nu$th Floquet state}, {\em i.e.}, such that $n_{\k,\nu'}=\delta_{\nu,\nu'}$. 
Then, the corresponding value of the pumped charge is:
\begin{equation} \label{eq:Q_F}
Q^{\rm F}_{\nu} = \frac{1}{3a \hbar\omega} \int_{0}^{\frac{2\pi}{a}} \ud k_y \; \frac{\partial \epsilon_{k_y,\nu}}{\partial k_y} \;,
\end{equation}
where we eliminated the trivial integral on $k_x\in [0,\frac{2\pi}{3a}]$.
To better understand the physical implications of this formula, let us start from the extreme adiabatic limit $\omega\to 0$,
where the predictions of the adiabatic theorem give us a $0$th-order expression for the
quasi-energies, in an extended Floquet BZ scheme~\cite{Holthaus_JPB16}, in the form:
\begin{eqnarray}\label{eq:qe_adiab}
\epsilon^{(0)}_{\k,\nu} &=& \frac{1}{\tau} \int_{\on}^{\tau} \ud t \left[ E_{\k,\nu}(t) - i\hbar \langle \phi_{\k,\nu}(t) | \partial_t \phi_{\k,\nu}(t) \rangle \right]  
\nonumber \\
&=& \epsilon^{d}_{\k,\nu} + \epsilon^{g}_{\k,\nu} \;.
\hspace{4mm}
\end{eqnarray}
Here $\phi_{\k,\nu}(t)$ and $E_{\k,\nu}(t)$ are the instantaneous eigenstates/eigenvalues of $\hat{H}(t)$, 
while $\epsilon^{d}_{\k,\nu}$ and $\epsilon^{g}_{\k,\nu}$ denote dynamical and geometric contributions~\cite{Niu_RMP10}. 
These contributions are in turn expressed as:
\begin{eqnarray}
\epsilon^{d}_{k_y,\nu} &=& a \int_{\on}^{\frac{2\pi}{a}} \!\! \frac{\ud k_x}{2\pi} \, E_{\k,\nu}(0) \nonumber \\
\epsilon^{g}_{k_y,\nu} &=& -\hbar\omega \int_{\on}^{\frac{2\pi}{a}} \!\! \frac{\ud k_x}{2\pi} \, \mathcal{A}_{x}^{(\nu)}(\k) \;.
\end{eqnarray}
%
where $\mathcal{A}_{x}^{(\nu)}(\k) =  i \langle \phi_{\k,\nu}(0) | \partial_{k_x} \phi_{\k,\nu}(0) \rangle $ is the Berry connection of the
$\nu$-th band. 
In both terms, the time integral has been transformed into a $k_x$-integral using the fact that the dependence on $t$ is
through the variable $ak_x+\omega t$. As a consequence, both terms are functions of $k_y$ only.  
The dynamical contribution is the $k_x$-averaged Bloch band, and is strictly periodic in $k_y$ of the BZ, see dashed red line 
in Fig.~\ref{fig:bands-chern}. On the contrary, the geometric term {\em winds} over the BZ, ending up acquiring an overall integer
equal to the Chern number of the corresponding band: 
\begin{equation}
\epsilon^{g}_{\frac{2\pi}{a},\nu} - \epsilon^{g}_{0,\nu} =  \hbar\omega C_\nu \;.
\end{equation}
This immediately leads to the expected integer quantization 
\begin{equation} \label{eq:Q_F_0}
3aQ^{(0)}_{\nu} = \frac{1}{\hbar\omega} \int_{0}^{\frac{2\pi}{a}} \ud k_y \; \frac{\partial \epsilon^{(0)}_{k_y,\nu}}{\partial k_y} =
C_{\nu} \;.
\end{equation}
%
\begin{figure}
\begin{center}
\def\svgwidth{90mm}
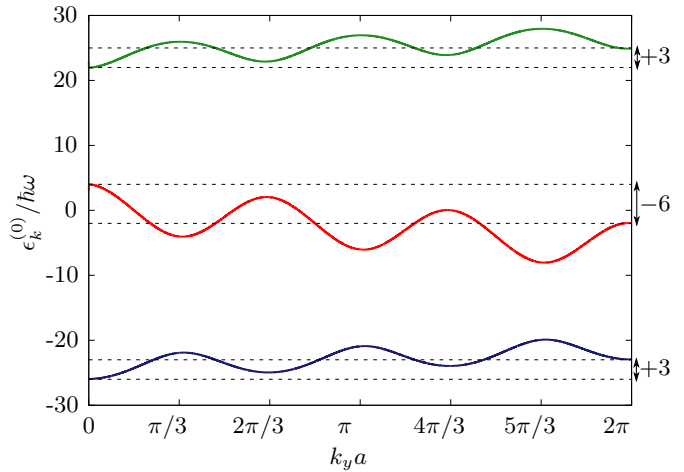
\caption{Adiabatic quasi-energies $\epsilon^{(0)}_{k_y,\nu}$ in units of $\hbar\omega$ for the three bands of the 
Harper-Hofsdtadter model. 
The frequency is $\hbar\omega=0.1J_0$.
$\epsilon^{(0)}_{k_y,\nu}$ is the sum of the $k_x$-averaged band, dashed line in Fig.~\ref{fig:bands-chern} (upper panel), plus 
the geometric contribution, giving rise 
to the loss of periodicity for $\epsilon^{(0)}_{k_y,\nu}$ in the BZ: $\epsilon^{(0)}_{\frac{2\pi}{a},\nu}=\epsilon^{(0)}_{0,\nu}+\hbar \omega C_\nu$, 
where $C_\nu$ is the Chern number of the $\nu$-the band.
}
\label{fig:adiab_bands}
\end{center}
\end{figure}

\begin{figure}[h!]
\begin{center}
\def\svgwidth{80mm}
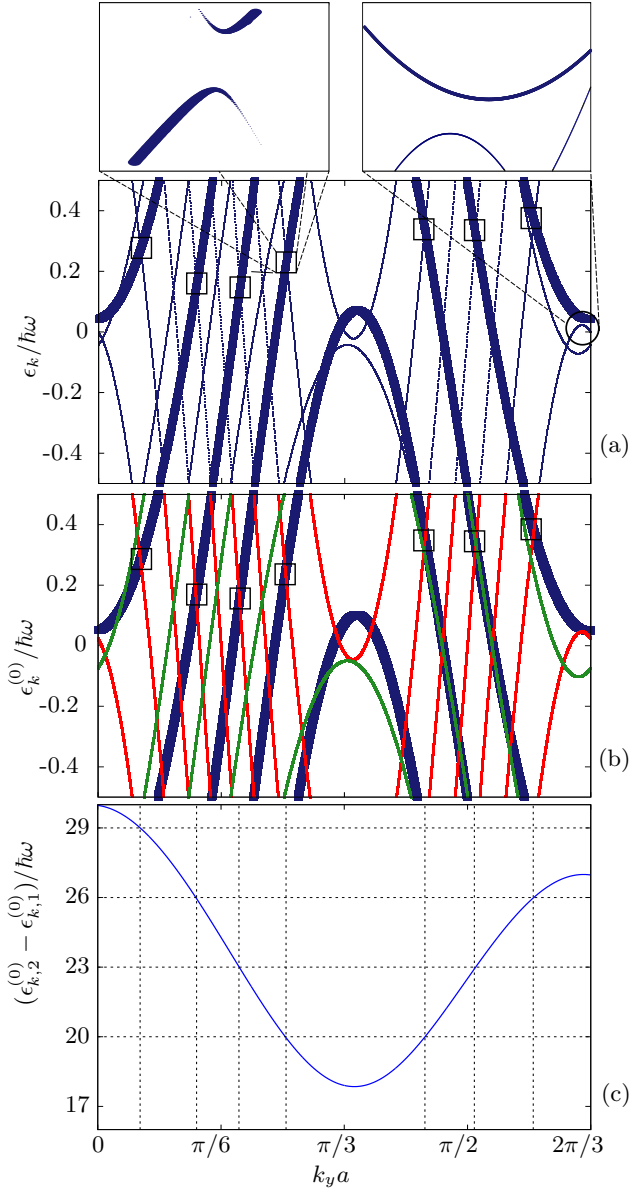
\caption{(a) Floquet spectrum as a function of $k_y$, for $\hbar\omega/J_0=0.1$. 
The squares signal the Floquet resonance avoided crossings, 
the circle an ordinary avoided crossing. Both are magnified in the top insets, where the size of the points is proportional to the 
$k_x$-averaged occupation $n_{k_y,\nu}$, see Eq.~\eqref{eq:nky_Floquet}. 
(b) Floquet adiabatic quasi energies, Eq.~\eqref{eq:qe_adiab}, folded in the Floquet BZ. 
(c) $(\epsilon^{(0)}_{k_y,2}- \epsilon^{(0)}_{k_y,1})/\hbar\omega$, the energy difference between the two lowest adiabatic bands 
in the extended-zone scheme, for $\hbar\omega=0.1J_0$. 
The vertical lines indicate the {\em Floquet resonances}, $\epsilon^{(0)}_{k_y,2}- \epsilon^{(0)}_{k_y,1}=m\hbar\omega$ with $m=29, 26, 23, 20$, 
giving rise to the avoided crossing gaps of panel (a). 
}
\label{fig:FL_eps}
\end{center}
\end{figure}

\begin{figure}
\begin{center}
\def\svgwidth{85mm}
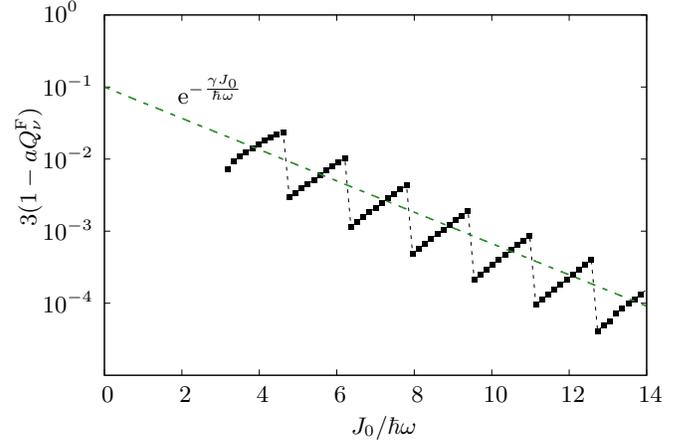
\caption{
The deviation of $3a Q_{\nu}^{\rm F}$, Eq.~\eqref{eq:Q_F}, from integer quantization, $3-3aQ_{\nu}^{\rm F}$, for the lowest Floquet band, assuming $n_{\k,\nu'}=\delta_{\nu,\nu'}$, 
showing that the exponentially small gaps in the quasi-energy spectrum at finite $\omega$ lead to exponentially small deviations.
 }
\label{fig:gaps}
\end{center}
\end{figure}

We now consider finite-$\omega$ effects beyond the adiabatic limit. 
Fig.~\ref{fig:FL_eps} shows the Floquet quasi-energy bands for $\hbar\omega/J_0=0.1$, plotted versus $k_y$ in the region
$[0,2\pi/(3a)]$, due to a periodicity $\epsilon_{k_y+\frac{2\pi}{3a},\nu}=\epsilon_{k_y,\nu}$ discussed in App.~\ref{app:Floquet}.
Notice that the quasi-energies are here naturally represented in the {\em Floquet Brillouin Zone}~\cite{Holthaus_JPB16} $[-\hbar\omega/2,\hbar\omega/2]$, 
as they are obtained by a numerical diagonalization of the Floquet operator. 
The thick line represents the quasi-energy band emerging from the low-energy band of Fig.~\ref{fig:adiab_bands}.
%
We observe two conspicuous features: 
\begin{description}
\item[i)] an apparent {\em winding} over the Floquet Brillouin Zone, as a quasi-energy crossing $\hbar\omega/2$ 
re-enters at $-\hbar\omega/2$ (and vice-versa). This is the winding expected from the geometric contribution to the adiabatic
quasi-energies shown in Fig.~\ref{fig:adiab_bands}. It would lead to:
\begin{equation}
\frac{1}{\hbar\omega} \int_{0}^{\frac{2\pi}{a}} \ud k_y \; \frac{\partial \epsilon_{k_y,\nu}}{\partial k_y} = C_{\nu} \;,
\end{equation}
where $C_\nu$ is the Chern number of the $\nu$-the band ($C_{\nu}=+3$, for the thick band shown in Fig.~\ref{fig:FL_eps}). 
\item[ii)] an apparent {\em crossing} of quasi-energies belonging to different Floquet bands. 
\end{description}
%
%
%
The crossings between different Floquet bands 
can develop very small {\em anti-crossing gaps} \cite{wigner_1929,Avron_JPA99},
as indeed we find at the points signalled by a square (see inset of Fig~\ref{fig:FL_eps}).  
To better understand the nature of such anti-crossing gaps, we reconsider again the adiabatic bands. 
The central panel of Fig~\ref{fig:FL_eps} shows a plot of the adiabatic bands $\epsilon^{(0)}_{k_y,\nu}$ folded back into the
Floquet BZ: quite evidently, they are a good approximation to the true quasi-energies for such value of $\omega$.
Notice, however, that here all the band crossings are genuine ones.
The bottom panel of Fig~\ref{fig:FL_eps}, finally, shows $\epsilon^{(0)}_{k_y,2}- \epsilon^{(0)}_{k_y,1}$, the energy difference between
the two lowest adiabatic bands, which clearly suggests that the anti-crossing points --- signalled by vertical dashed lines ---
are associated to {\em Floquet resonances} when $\epsilon^{(0)}_{k_y,2}- \epsilon^{(0)}_{k_y,1}=m\hbar\omega$. 
Surprisingly, not all possible resonances actually lead to the opening
of an anti-crossing gap, but only a sequence of them, here with $m=29, 26, 23, 20$. 
The periodicity of $\Delta m=3$ is likely associated to our choice of flux $\alpha=1/3$, but the precise location of the resonance 
is not fully understood. One thing that we can say, however, is that the resonances open up gaps~\cite{Shih_PRB94} in the quasi-energy spectrum 
that are {\em exponentially small in $1/\omega$}.
This makes such gaps quite difficult to pin-point precisely, but our numerical evidence is reasonably robust on that issue.
%
%
Fig.~\ref{fig:gaps} shows the deviation from integer quantization, $3-3aQ_{\nu}^{\rm F}$,
--- calculated assuming $n_{\k,\nu}=1$ and using Eq.~\eqref{eq:Q_diag_orig}, which, as opposed to Eq.~\eqref{eq:Q_F}, avoids 
derivatives of numerically determined quasi-energies --- as a function of $J_0/\hbar\omega$: in the $\omega$-region we plot, 
an overall exponential decay is clearly visible for $3-3aQ_{\nu}^{\rm F}\sim \nep^{-\gamma J_0/(\hbar\omega)}$,
with $\gamma\sim 0.5$,  
superimposed on a saw-tooth behaviour due to the sudden formation of larger gaps when two nearby gaps coalesce together upon decreasing
$\omega$. 
%
%
Summarizing, if the quasi-energy avoided-crossing gaps opening were the main responsible for finite-frequency/field corrections 
to the quantized pumped charge, such deviations would be exponentially small in $1/\omega\propto 1/F$: 
therefore non-analytic in the field strength \cite{Avron_JPA99} and exceedingly small for most practical purposes:
for instance, in an experiment in which $\hbar\omega=10^{-2} J_0$, we would estimate $3-3aQ_{\nu}^{\rm F}\approx 10^{-22}$. 


\subsection{Effect of the occupation factors: sudden switch-on.}\label{sseq:quench}
\begin{figure}
\begin{center}
\def\svgwidth{85mm}
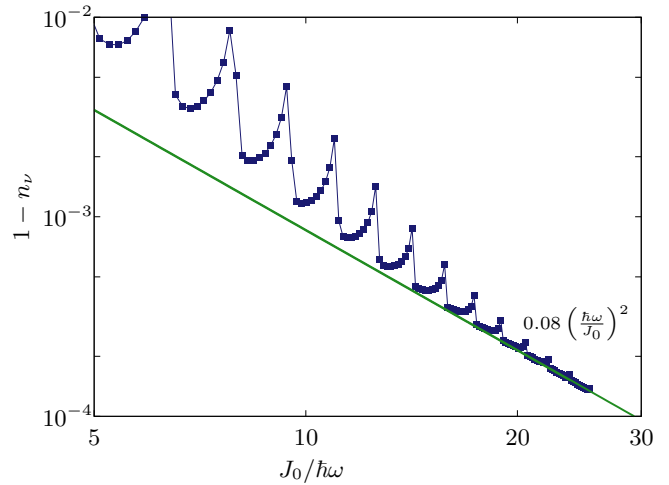
\caption{Correction to the $\k$-averaged adiabatic Floquet mode occupation $n_{\nu}$, Eq.~\eqref{eq:nad}, vs $1/\omega$, 
showing the good agreement between the numerical data and the perturbation theory prediction from Eq.~\eqref{eq:APTcorr}.
}
\label{fig:nad}
\end{center}
\end{figure}

\begin{figure}
\begin{center}
\def\svgwidth{85mm}
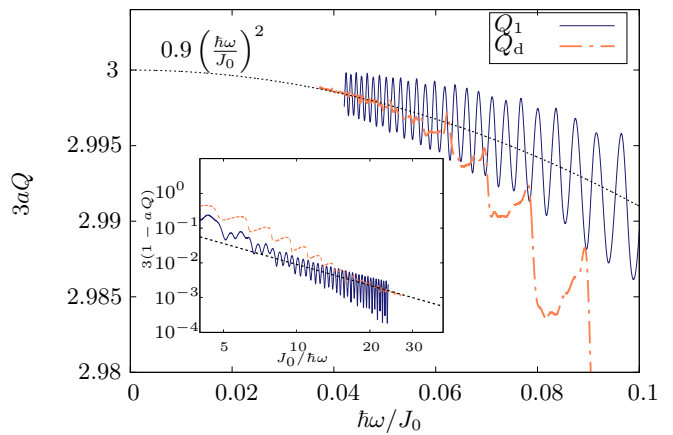
\caption{Pumped charge vs $\omega$, both for one period (solid line) $3aQ_1$, Eq.~\eqref{eq:Q_1}, and in the diagonal ensemble (dashed line) 
$3aQ_{\rm d}$, Eq.~\eqref{eq:Q_diag_bis}. 
Notice the oscillations in $Q_1$, signalling an essential singularity in $\omega=0$. 
The inset shows the deviation from the quantized value $3aQ_1(\omega \rightarrow 0)=3$ vs $1/\omega$. 
For small frequency this deviation is quadratic in $\omega$.
}
\label{fig:Qquench}
\end{center}
\end{figure}

The second source of deviations from perfect quantization arises from the fact that the prepared state $|\psi(t_{\on})\rangle$ is 
{\em not precisely a Floquet state}, i.e., that Floquet occupation factors deviate from $n_{\k,\nu'}=\delta_{\nu,\nu'}$. 
The inset of Fig.~\ref{fig:FL_eps}, where the size of the dots is proportional to the Floquet occupation, shows that sizeable
deviations occur whenever $\omega>0$, even if small, at the quasi-energy avoided level crossing.  
Indeed, for a quasi-adiabatic evolution, the Floquet modes will be ``close'' to the eigenstates of the Hamiltonian, to which they reduce 
for $\omega\rightarrow 0$. 
If we initialize the system in an insulating phase by filling the lowest-energy band,
one of the Floquet occupation number $n_{\k,\nu}$ will be close to $1$ and much higher than the others: 
the corresponding Floquet mode will be the main one responsible for charge transport. 
In the following, we will refer to such a state as {\em adiabatic} or {\em lowest-energy} Floquet state:
it is indeed the Floquet state which has the largest overlap with the instantaneous Hamiltonian ground state.
This is highlighted in the Fig.~\ref{fig:FL_eps}, where the Floquet spectrum is plotted vs $k_y$
with thickness proportional to $k_x$-averaged occupation factor
\begin{equation} \label{eq:nky_Floquet}
n_{k_y,\nu} = \frac{3a}{2\pi} \int_{0}^{\frac{2\pi}{3a}} \!\! \ud k_x \; n_{\k,\nu} \;.  
\end{equation}
%
%
%
Let us now focus on the occupation of such ``adiabatic'' Floquet state. 
If the driving field is suddenly turned on from $F_x(t\le 0)=0$ to $F_x(t>0)=F$, $|\psi(t_{\on}=0)\rangle$ coincides with a Slater determinant Bloch
eigenstate of $\hat{H}(0)$, and $n_{\k,\nu}$ is given by the overlap of such a state with the adiabatic Floquet state: 
$n_{\k,\nu}=| \langle u_{\k,\nu}(0) | \phi_{\k,\nu}\rangle |^2$.
When $\omega$ is small, we can combine adiabatic perturbation theory\cite{Rigolin_PRA08} (APT) 
to obtain an approximate expression for the Floquet modes $|u_{\k,\nu}(0)\rangle$, see App.~\ref{app:pt} for details. 
Following this approach $n_{\k,\nu}$ can be calculated to be:
\begin{equation}\label{eq:APTcorr}
n_{\k,\nu} = 1-\left(\frac{\hbar\omega}{2\pi}\right)^2 \sum_{\mu \neq \nu} 
\bigg| \frac{M_{\mu,\nu}^{(\k)}}{\Delta_{\mu,\nu}^{(\k)}} \bigg| ^2 + O(\omega^3) \;.
\end{equation}
Here $M_{\mu,\nu}^{(\k)}$ and $\Delta_{\mu,\nu}^{(\k)}$ are calculated from instantaneous Hamiltonian eigenvalues/eigenstates,
$\hat{H}_{\k}(s)|\phi_{\k,\nu}(s)\rangle=E_{\k,\nu}(s) |\phi_{\k,\nu}(s)\rangle$ where $s=\omega t$ 
is the rescaled time, as:  
\begin{eqnarray}
\Delta_{\mu,\nu}^{(\k)}(s) &=& E_{\k,\mu}(s)-E_{\k,\nu}(s) \;, \nonumber \\
M_{\mu,\nu}^{(\k)}(s)     &=& \frac{\langle \phi_{\k,\mu}(s)|\partial_s \hat{H}_{\k}(s)| \phi_{\k,\nu}(s)\rangle}
{\Delta_{\nu,\mu}^{(\k)}} \;.
\end{eqnarray}
In Eq.~\eqref{eq:APTcorr} all quantities are evaluated at $s=2\pi$, corresponding to $t=\tau$, a full period.
Therefore, if the matrix elements $M_{\mu,\nu}$ are not all equal to zero, which in general they are not, we expect to see power-law 
corrections to the occupation number of the Floquet states, leading to a similar behaviour for the pumped charge. 
Fig.~\ref{fig:nad} shows the $\k$-averaged occupation 
\begin{equation} \label{eq:nad}
n_{\nu}=3a^2 \int_{\rm BZ} \frac{\ud^2 \k}{(2\pi)^2} \; n_{\k,\nu} \;,   
\end{equation}
calculated numerically, compared to the perturbation theory estimate in Eq.~\eqref{eq:APTcorr}, as a function of $\omega$: 
the $\omega^2$ deviation is quite clearly visible.
This quadratic correction to the occupation factors reflects itself into the pumped charge, both the single-period charge $Q_1$,
Eq.~\eqref{eq:Q_1}, 
as well as the infinite-time average $Q_{\rm d}$, Eq.~\eqref{eq:Q_diag_bis}, as seen from Fig.~\ref{fig:Qquench}. 
The faster-and-faster oscillations seen in $Q_1$ for $\omega\to 0$ originate from the essential singularity in $\omega=0$ 
of the expectation value of the current operator \cite{Avron_JPA99}; the oscillations are smeared in $Q_{\rm d}$, due to the infinite-time average.
This behaviour is very similar to that reported in Ref.~\onlinecite{Privitera_PRL18} for adiabatic quantum pumping in the
Rice-Mele model.

\subsection{Effect of the occupation factors: continuous switch-on.}\label{sseq:ramp}
\begin{figure}
\begin{center}
\def\svgwidth{85mm}
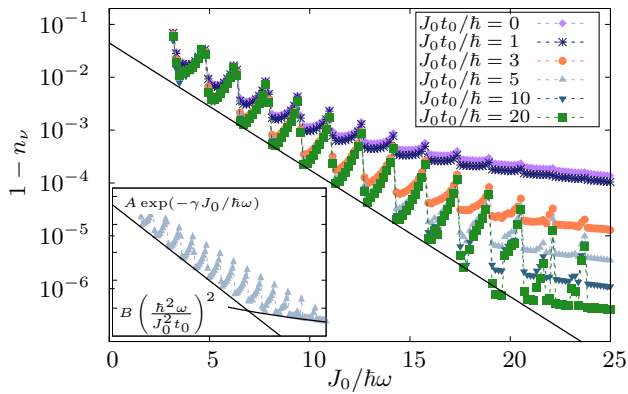
\caption{
Correction to the occupation of the lowest energy Floquet band for different switch-on times $t_{\on}$. 
The inset highlights the crossover form an exponential regime $\nep^{-1/\omega}$ to the quadratic one $\omega^2$ for finite ramp time $t_{\on}$. 
The solid lines correspond to the functions $0.05 \nep^{-\frac{0.5 J_0}{\hbar \omega}}$ and $0.002 (\hbar\omega/J_0)^2$.}
\label{fig:Qrampvsquench}
\end{center}
\end{figure}

The picture becomes richer if we switch-on the driving in a continuous fashion, taking $F_x(t)=Ff(t/t_{\on})$ with a suitably 
smooth function $f(s=t/t_{\on})$.
The first obvious choice is a linear switch-on, $f(s)=s$, with a fixed switch-on time $t_{\on}$.
As shown in Fig.~\ref{fig:Qrampvsquench}, we now observe two regimes: 
a first one, for relatively large $\omega$, where the corrections to the occupation $n_{\nu}$ of the adiabatic Floquet band appear to be
exponentially small in $1/\omega$, and a second regime, for small $\omega$, where the corrections are $\propto \omega^2$:
\begin{equation}
1-n_{\nu} \sim \left\{ 
\begin{array}{ll}
A \, \displaystyle \nep^{-\frac{\gamma J_0}{\hbar \omega}} & \;\; \mbox{for} \;\; \omega > \omega^*  \vspace{3mm} \\
B \, \displaystyle \frac{\hbar^4\omega^2}{J_0^4 t_{\on}^2}  & \;\; \mbox{for} \;\; \omega < \omega^* 
\end{array}
\right. \;.
\end{equation}
The two regimes have markedly different behaviours. 
The non-analytic exponential observed at higher $\omega$ is {\em universal} 
--- with $\gamma \simeq 0.5$ from our data, and at most a very mild dependence of $A$ on $t_{\on}$ --- 
and, as we will argue, it is directly related to the width of the resonances of the Floquet spectrum.
The power-law regime is non-universal, with an amplitude decreasing as $1/t_{\on}^2$: hence the crossover frequency 
$\omega^*$ between these two regimes, which is approximately given by:
\begin{equation}
\frac{J_0}{\hbar\omega^*} \simeq \log \left(\sqrt{\frac{A}{B}} \frac{J_0 t_{\on}}{\hbar}\right)\ ,
\end{equation}
is shifted towards smaller $\omega$ as $t_{\on}$ increases.
%
Notice that the crossover $\omega^*$ exists only if $\frac{J_0t_{\on}}{\hbar}\ge \sqrt{\frac{B}{A}}\frac{\nep}{4}$; 
indeed if $t_{\on}$ is too small, only the power law regime survives, leading to the ordinary ``perturbative response'' 
observed for the sudden quench case.
%
\begin{figure}
\begin{center}
\def\svgwidth{85mm}
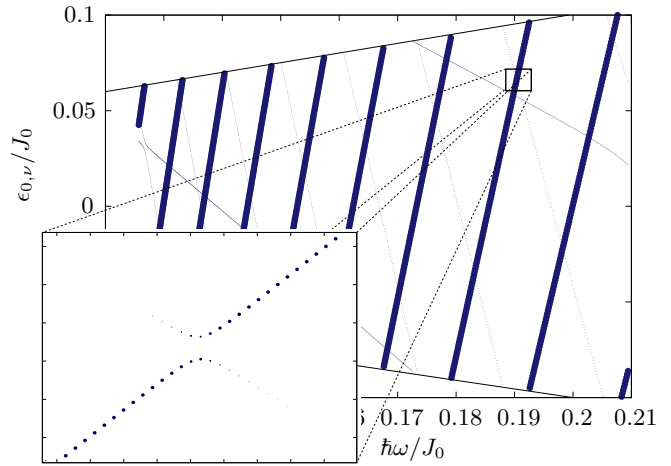
\caption{
Floquet quasi-energies in ${\bf k}=0$ as a function of the frequency. 
The width of the line is proportional to the occupation of the state when the system is in the ground state with filling factor $1/3$. 
The inset zooms on a level crossing to highlight the presence of gaps, showing also that the adiabatic band is the ``excited'' state 
after the avoided crossing.
The solid black lines are the boundary of the first Floquet-Brillouine zone.
}\label{fig:floqgaps}
\end{center}
\end{figure}

It is interesting to ask why the the continuity in time of the force field $F_x(t)$ is so important. 
As explained in Sec.~\ref{sseq:floq} the topological properties at finite frequency are related to the Floquet states,
while the system is initially prepared in a state $|\psi(0)\rangle$ which coincides with the Hamiltonian ground state. 
By switching on the driving force in a continuous manner, $F_x(t)=F f(t/t_{\on})$, the initial state is {\em continuously} 
deformed into a state which is ``closer'' to the ``lowest-energy'' Floquet state at the final frequency $\omega$. 
Fig.~\ref{fig:floqgaps} helps to illustrate what happens as we turn on the driving frequency: as the instantaneous $\omega(t)=aF_x(t)/\hbar$ 
increases, each Floquet mode winds around the expanding Floquet-Brillouine zone (FBZ) and encounters a series of (avoided) 
level crossings in the quasi-energy spectrum, with exponentially small gaps $\Delta$. 
Since the gaps $\Delta$ are exponentially small, however, a finite value of $t_{\on}$ will lead the system to cross them {\em diabatically}.
The final Floquet state will show an occupation which can be interpreted \cite{Breuer_PLA89} 
as the excitation probability after many Landau-Zener\cite{Zener696,Landau_qmech:book} events.
Following Ref.~\onlinecite{Breuer_PLA89}, at each avoided crossing we obtain a transition probability 
\begin{equation}\label{eq:LZ}
P_{\rm ex}(\omega,t_{\on})=\nep^{-\frac{\Delta^2 t_{\on}}{4\hbar^2 \zeta \omega}} \,,
\end{equation}
where we used the fact that the speed at which the gap is crossed can be estimated as 
$\partial_t(\epsilon_2-\epsilon_1) \simeq \zeta \hbar \omega/t_{\on}$,
$\zeta$ being the difference in slope between the two quasi-energy bands as they wind around the Floquet BZ. 
Since the gaps $\Delta (\omega)\sim J_0 \nep^{-\gamma J_0/ \hbar\omega}$ are the smallest quantities, it is legitimate to
expand the exponential in Eq.~\eqref{eq:LZ} to lowest order in $\Delta^2$. 
A further simplification is due to the fact that the dominant contribution to the sequence of LZ processes comes 
from the largest gap encountered, which correspond to the end of the ramp, when the frequency is maximum.
Hence we obtain the following estimate for the corrections to $n_\nu$
\begin{equation}\label{eq:occ_cor}
1-n_\nu \sim \frac{\Delta^2 t_{\on}}{4\hbar^2 \zeta \omega} \sim 
\frac{J_0^2 t_{\on}}{4\hbar^2 \zeta\omega} \, \nep^{-2\gamma J_0/\hbar\omega} \;.
\end{equation}
This rather crude estimate gives a hint on the physical mechanism behind the non-perturbative corrections to the integer 
occupation of the Floquet mode
observed when the electric field is turned on at a finite rate $1/t_{\on}$.
Incidentally, Eq.~\eqref{eq:occ_cor} also suggests that increasing the ramp time $t_{\on}$ would lead to {\em larger corrections} 
to both $n_\nu$ and $Q_{\rm d}$, although our numerical data do not show this, possibly because of the limited range of  
 $t_{\on}$ explored. 
Indeed, at the level crossing, a larger $t_{\on}$ would increase the adiabaticity of the process, 
therefore decreasing the occupation of the lowest energy Floquet mode, which corresponds to the ``excited'' state in the quasi-energy spectrum, 
as shown in Fig.~\ref{fig:floqgaps}.	
An alternative explanation is given in Ref.~\onlinecite{Polkovnikov_PhysRep17}, 
where it is suggested that the increasing deviations from the adiabatic preparation of Floquet states for very large ramp times $t_{\on}$ 
is related to the absorption of energy from the external field, leading to heating of the system.

\begin{figure}
\begin{center}
\def\svgwidth{85mm}
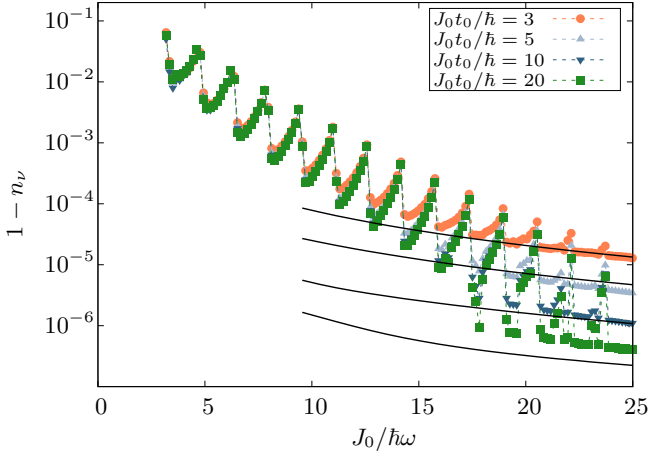
\caption{Correction to the occupation of the $\nu-$th Floquet mode compared with the prediction of Eq.~\eqref{eq:Floq:ad_apr} (solid black lines).}\label{fig:floq_apt}
\end{center}
\end{figure}
This picture breaks down for small $\omega$, where the crossover with the quadratic regime occurs.
The observed $\omega^2$ scaling suggests that a Floquet adiabatic perturbation theory (FAPT)\cite{breuer1989adiabatic,Polkovnikov_PhysRep17,Drese_EPJD99} might be appropriate here.  
Unfortunately, the standard framework of application of such a theory is when the slowly changed parameters $\lambda(t)$
do not involve the crossing of Floquet resonances \cite{Polkovnikov_PhysRep17}, which is certainly not the case for $\omega\to 0$.
So, we construct here a simplified version of FAPT which should capture the $\omega\to 0$ regime. 
To do so, we start from an expansion of the state $|\psi_{\k}(t)\rangle$ in terms of 
instantaneous Floquet modes $|u_{\k,\mu}(\omega(t),t)\rangle$ corresponding to a frequency $\omega(t)$ (which is slowly evolving in time),
with associated phase factor given by the {\em adiabatic} Floquet quasi-energy $\epsilon^{(0)}_{\k,\mu}(\omega(t))$:
\begin{equation}
|\psi_{\k}(t)\rangle = \sum_{\mu} c_{\k,\mu}(t) \nep^{-\frac{i}{\hbar} \int_0^t \epsilon^{(0)}_{\k,\mu}(\omega(t'))} |u_{\k,\mu}(\omega(t),t)\rangle \;.
\end{equation}
Proceeding as in the standard APT, assuming that at $t=0$ we have $c_{\k,\mu}(0)=\delta_{\mu,\nu}$ and keeping only the 
lowest-order terms we end-up writing:
\begin{equation} 
c_{\k, \mu \neq \nu} (t_{\on}) 
\approx - \int_0^{\omega} \!\ud \omega' \, \langle u_{\k,\mu} | \partial_{\omega'} u_{\k,\nu} \rangle
\nep^{-\frac{it_{\on}}{\hbar\omega} \int_0^{\omega'} (\epsilon^{(0)}_{\k,\nu}-\epsilon^{(0)}_{\k,\mu})}
\nonumber
\end{equation}
where we assumed a linear adiabatic switch-on, $\omega(t)=(t/t_{\on}) \, \omega$, and changed variable to an integral over frequency.
Here $|u_{\k,\mu}\rangle$ stands for $|u_{\k,\mu}(\omega',t(\omega')\rangle$, where $t(\omega')=t_{\on} \omega'/\omega$.
Noticing now that the adiabatic quasi-energy differences $(\epsilon^{(0)}_{\k,\nu}-\epsilon^{(0)}_{\k,\mu})$ are large compared to $\omega'$,
we integrate by part, as in standard APT, ending up with:
\begin{widetext}
\begin{equation}\label{eq:Floq:ad_apr}
c_{\k, \mu \neq \nu} (t_{\on}) \approx \frac{i\hbar\omega}{t_{\on}} 	
\frac{\langle u_{\k,\mu}(\omega',t(\omega')) | \partial_{\omega'} u_{\k,\nu}(\omega',t(\omega')) \rangle}
{\epsilon^{(0)}_{\k,\nu}(\omega')-\epsilon^{(0)}_{\k,\mu}(\omega')} 
\nep^{-\frac{it_{\on}}{\hbar\omega} \int_0^{\omega'} (\epsilon^{(0)}_{\k,\nu}-\epsilon^{(0)}_{\k,\mu})} \bigg|_{\omega'=0}^{\omega'=\omega} \;.
\end{equation}
\end{widetext}
Finally we compute the scalar products $\langle u_{\k,\mu} | \partial_{\omega'} u_{\k,\nu} \rangle$ by using the expansion derived in App.~\ref{app:pt},
in particular Eq.~\ref{eq:corr1}, which allows us to write:
\begin{equation} \label{eqn:scalar_product}
\langle u_{\k,\mu} | \partial_{\omega'} u_{\k,\nu} \rangle=
\frac{M_{\mu,\nu}^{(\k)}}{\Delta_{\mu,\nu}^{(\k)}}+O(\omega')\;.
\end{equation}
Substituting back into Eq.~\eqref{eq:Floq:ad_apr}, we get an expression that can be computed numerically.
%
%
%
%
Once the projections $c_{\k,\mu \ne \nu}$ have been computed, the correction to the occupation number of the ``adiabatic'' Floquet state reads
\begin{equation}
1-n_{\nu}=3 a^2 \sum_{\mu \ne \nu} \int_{BZ} \frac{\ud^2 {\bf k}}{(2\pi)^2} |c_{\k,\mu}|^2 \;.
\end{equation}
As shown in Fig.~\ref{fig:floq_apt} this simplified FAPT describes quite well the quadratic regime and its scaling with $t_0$.
We observe that the accuracy of the approximation seems to decrease as $t_0$ grows, probably because non-adiabatic corrections 
to the time-evolved eigenstates need to be taken into account in computing Eq.~\eqref{eqn:scalar_product}. 

\begin{figure}
\begin{center}
\def\svgwidth{85mm}
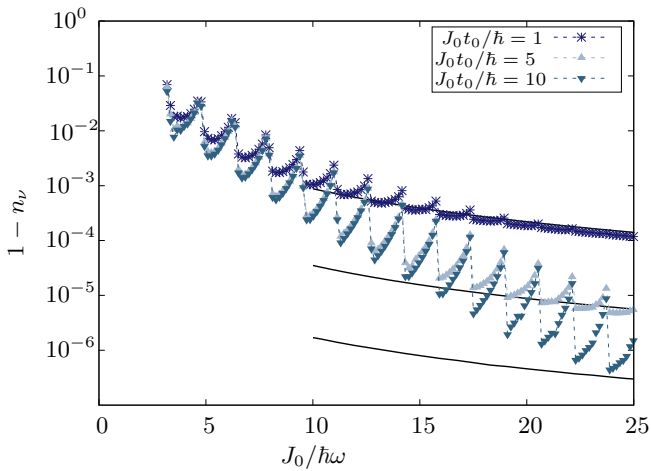
\caption{
Correction to the occupation number of the adiabatic Floquet state, when the driving force is smoothly turned on with the switching function 
$f(s=t/t_{\on})=\frac{1}{2} \left( 1 - \cos \left( \pi s \right) \right)$.
Eq.~\eqref{eq:Floq:ad_apr}, with a slight modification due to the different driving schedule, still gives a good estimate of the quadratic regime for 
small $\omega$.}
\label{fig:occSramp}
\end{center}
\end{figure}

As a final check, we have considered whether imposing continuity also on the first derivative of $\omega(t)$ makes any difference or not. 
Fig.~\ref{fig:occSramp} shows the occupation of the ``adiabatic'' Floquet state when the frequency is increased smoothly from 0 to its final value 
$\omega$ with a switching function $f(s) = \frac{1}{2} \left( 1 - \cos \left( \pi s \right) \right)$.
Beside some small numerical difference, the situation is qualitatively similar to the one obtained with the linear ramp (Fig.~\ref{fig:Qrampvsquench}), with a crossover between an exponential regime for $\omega > \omega^*(t_{\on})$ and a power law tail for $\omega < \omega^*(t_{\on})$.
This suggests that while a {\em necessary} condition --- albeit not sufficient --- to obtain non-perturbative corrections is indeed the {\em continuity} of 
the force field $F_x(t)$, its differentiability seems not to be required. 

%
%
%
\section{Conclusions}\label{sec:conclusion}
In this paper we investigated the robustness of the quantization of the Hall conductivity beyond the validity range of linear response theory (adiabatic limit), 
in the Harper-Hofstadter model.
This work was mainly motivated by the possibility of realizing simple tight-binding Hamiltonians, such as the HH one, in cold atoms experiments with 
synthetic gauge fields, where the model parameters can be easily fine-tuned.
By employing Floquet theory for time-periodic systems, we showed that the quantization of the transverse pumped charge $Q_{\rm d}$  depends mainly on the occupation factor $n_{\nu}$ of the lowest energy Floquet state.
In particular, we found that a continuous and sufficiently slow switching-on of the driving force is necessary to obtain corrections to the Kubo formula which
are non-analytic in the force amplitude $F$, scaling as $\nep^{-\gamma/|F|}$.
If the switching time $t_{\on}$ is too small, or the force is turned on abruptly, corrections of the order $O(F^2)$ are always recovered when $F\to 0$.
A crossover force amplitude $F^*(t_\on)$ between the quadratic and the exponential regimes is clearly shown by our numerical analysis for any finite switching time 
$t_\on$, and it would be interesting to see if this crossover can be indeed be observed in experimental realizations of IQHE or quantum pumping in 
optical lattices experiments.

Future investigations could focus on how the robustness of the topological phase and the crossover with a perturbative regime are affected by the presence 
of disorder or dissipation.
With regard to disorder, it is well known that in solid state realizations of IQHE a certain amount of impurities, with associated localized states, are 
crucial to the robustness of the Hall plateaus. 
The robustness of the topological state against disorder~\cite{Zhang_ChPhysB13} or absence of translational invariance~\cite{Puschmann_arx18} 
has also been tested in simple tight-binding models: the crucial question, for what concerns our story, is if disorder tends to increase the ``robustness'' of 
the time response, as we have formulated it, by increasing the extent of the region in which non-analytic corrections to the Kubo formula dominate. 
%
We observe that the dimensionality might play a role: while for clean samples a two-dimensional (2D) lattice model with a constant drift is essentially equivalent 
to a one-dimensional (1D) chain with a time periodic driving, such as the Rice-Mele model \cite{Privitera_PRL18}, disorder could affect 1D and 2D system 
in different ways.

Similar questions can be formulated concerning the role of dissipation: while the linear response regime is quite well understood~\cite{Ozawa_PRL14,Ozawa_PRB18}, 
the interplay between non-adiabatic effects and the coupling with a thermal bath still requires a precise characterization.
Preliminary results on the effect of dissipation in the periodically driven Rice-Mele model \cite{Arceci:unpub} show that dissipation towards a low-temperature bath
can be beneficial in increasing the occupation of the lowest-energy Floquet states, thus making the pumped charge closer to the Thouless adiabatic limit.

\section*{Aknowledgments}
We wish to thank L. Privitera, S. Fazio, and A. Russomanno for useful discussions.
GES acknowledges support by the EU FP7 under ERC-MODPHYSFRICT, Grant Agreement No. 320796.

\appendix
\section{Adiabatic expansion of Floquet eigenstates} \label{app:pt}
In this appendix we derive an expansion in powers of $1/\tau$ for the Floquet modes, looking in particular for their overlap 
with the Hamiltonian eigenstates. 
To obtain this expansion, we first exploit adiabatic perturbation theory\cite{Rigolin_PRA08} (APT) to compute the Floquet operator 
and then we use ordinary perturbation theory (PT) to calculate the corrections to the Floquet modes. 

Given a gapped periodic Hamitlonian $\hat{H}(t+\tau)=\hat{H}(t)$, 
and $\hat{H}(t)|\phi_\nu(t) \rangle=E_\nu(t)|\phi_\nu(t) \rangle$ denotes instantaneous eigenstates/eigenvalues, 
the adiabatic theorem states that, if the evolution is slow enough, we can write the 
{\em time evolved} state $|\psi_{\nu}(\tau)\rangle$ originating from 
$|\psi_{\nu}(t=0)\rangle \equiv |\phi_{\nu}(0)\rangle$, to $0-$th order in $1/\tau$, as: 
\begin{equation}
|\psi_{\nu}(\tau)\rangle \approx |\psi_{\nu}^{(0)}(\tau)\rangle = \nep ^{-i \epsilon_\nu \tau/\hbar} |\phi_\nu(0)\rangle \;,
\end{equation}
where  {
\[ \epsilon_{\nu} \tau= \int_{0}^\tau \! \ud t \; \big( E_{\nu}(t)-i\hbar \langle \phi_\nu(t)|\partial_t \phi_{\nu}(t)\rangle \big) \;. \]
Hence, writing the Floquet operator as
\begin{equation}\label{eq:floq_generic}
\hat{F}(\tau)\equiv \hat{U}(\tau,0) = \sum_\nu |\psi_\nu (\tau) \rangle \langle \phi_\nu(0) | \;,
\end{equation}
the adiabatic theorem tells us that it reduces when $\omega=\frac{2\pi}{\tau}\rightarrow 0$ to the expression
\begin{equation}\label{eq:floq_adia}
\hat{F}^{(0)}(\tau)=\sum_\nu \nep ^{-i \epsilon_\nu \tau/\hbar }|\phi_\nu(0) \rangle \langle \phi_\nu(0) | \ .
\end{equation}
Notice that $|\phi_\nu  (\tau) \rangle =|\phi_\nu(0) \rangle $ due to the time periodicity of the Hamiltonian. 
Eq.~\eqref{eq:floq_adia} means that in this limit the Floquet modes $|u_\nu(0)\rangle$ and the instantaneous eigenstates $|	\phi_\nu(0)\rangle$ at the end of each period coincide.  
To obtain finite frequency corrections, we need to write $|\psi_\nu (\tau) \rangle$ through an adiabatic perturbation series\cite{Rigolin_PRA08}
\begin{equation}
|\psi_\nu(\tau) \rangle =\sum_{p=0}^{\infty} \left( \frac{\hbar}{\tau}\right)^p|\psi^{(p)}(\tau)\rangle \;, 
\end{equation}
which leads to a similar expression for the Floquet operator
\begin{equation}
\hat{F}(\tau) = \sum_{p=0}^{\infty} \left( \frac{\hbar}{\tau} \right)^p\hat{F}^{(p)}(\tau) \;.
\end{equation}
Now we assume that at $t=0$ the system is prepared in the  $\nu-$th eigenstate $|\phi_\nu(0)\rangle $ of the Hamiltonian.
From Eq.~\eqref{eq:floq_adia} we expect a single Floquet state to have a large overlap with $|\phi_\nu(0)\rangle $
if the period $\tau$ is large,  and therefore its occupation number $n_\nu$ should be close to one.
We wish to exploit perturbation theory to compute the lowest order corrections in $\frac{1}{\tau}$,	to
%
\begin{equation}\label{eq:app_occ}
n_{\nu}=\left| \langle \phi_\nu(0) | u_{\nu}(0)\rangle \right|^2 = \left| \langle \phi_\nu(\tau) | u_{\nu}(\tau)\rangle \right|^2  \;.
\end{equation}
As we will show in the following, the lowest order terms are quadratic in $\frac{1}{\tau}$ ---or equivalently in $\omega$--- and they originate from second order corrections due to $\hat{F}^{(1)}(\tau)$.
Indeed first order corrections to a given eigenstate in perturbation theory are always orthogonal to the unperturbed one
and therefore $\hat{F}^{(2)}(\tau)$ can only give contribution of order $O(\frac{1}{\tau^4})$.
Thus we just neep to compute $\hat{F}^{(1)}(\tau)$.
Before proceeding we define the following quantities depending on a rescaled time $s=t/\tau$
\begin{eqnarray*}
\tau \epsilon_\nu &=& \tau \int_{\on}^1 \ud s E_\nu(s) -i\hbar \int_{0}^1 \ud s \langle \phi_\nu(s) | \partial_s \phi_\nu(s) \rangle \ , \\ 
\Delta_{\nu, \mu}(s) &=& E_{\nu}(s)-E_{\mu}(s) \ ,
\end{eqnarray*}
where $\lbrace E_\nu(s) \rbrace$ is the set of instantaneous eigenvalues of $\hat{H}(s)$ and $\lbrace |\phi_\nu (s) \rangle \rbrace$ the corresponding eigenvectors.
The adiabatic expansion of the time evolved state $\psi_\nu(t)$ will be written in terms of 
\begin{eqnarray*}
M_{\mu,\nu}(s) &= 
& \frac{\langle \phi_{\mu}(s)|\partial_s \hat{H}(s)| \phi_{\nu}(s)\rangle}{\Delta_{\nu,\mu}} \ ,\\
J_{\mu,\nu} &=& \int_{\on}^1 \ud s \frac{|M_{\mu,\nu}(s)|^2}{\Delta_{\mu,\nu}(s)} \ ,
\end{eqnarray*}
which again depend only on the instantaneous spectrum of the Hamiltonian.
Following Ref.~\onlinecite{Rigolin_PRA08} we can write the first-order correction to the evolved eigenstate $|\psi_{\nu}(\tau)\rangle$  as: 
\begin{equation}\label{eq:appAPT}
\begin{split}
|\psi^{(1)}_\nu\rangle = & i \sum_{\mu \neq \nu} \nep^{-i\tau\epsilon_{\nu}/\hbar} J_{\mu,\nu} |\phi_{\nu} \rangle 
+ \\
 &i \sum_{\mu \neq \nu} \frac{M_{\mu,\nu}}{\Delta_{\mu,\nu}} \left(\nep^{-i\tau\epsilon_\nu/\hbar} - \nep^{-i\tau\epsilon_\mu/\hbar} \right) |\phi_\mu\rangle \ ,
\end{split}
\end{equation}
 where the $s$ dependence is omitted since all quantities are computed in $s=1$.
The ``perturbation'' of order $1/\tau$ to the Floquet operator consists in a diagonal part (first term of the RHS in Eq.~\eqref{eq:appAPT}) and an off-diagonal part (second term).
The former acts only as a renormalization of the eigenvalues (the Floquet quasi-energies) of the operator but does not change the eigenvector, since it is diagonal in the original basis.
We can now apply perturbation theory for linear operators to obtain the correction to the Floquet modes.
The first order term reads
\begin{equation} \label{eq:corr1}
\begin{split}
|u_\nu^{(1)}\rangle &= 	\frac{\hbar}{\tau}	\sum_{\mu \neq \nu} |\phi_\mu \rangle \frac{\langle \phi_\mu| \hat{F}^{(1)}(\tau)|\phi_\nu\rangle}{\nep^{-i\tau\epsilon_\nu/\hbar}-\nep^{-i\tau\epsilon_\mu/\hbar}} \\
& = 
i \frac{\hbar}{\tau} \sum_{\mu \neq \nu} \frac{M_{\mu,\nu}}{\Delta_{\mu,\nu}} |\phi_\mu\rangle \ ,
\end{split}
\end{equation}
where the off diagonal elements of $F^{(1)}(\tau)$ are obtained by combining Eq.~\eqref{eq:floq_generic} and Eq.~\eqref{eq:appAPT}, leading to
\begin{equation}
F^{(1)}_{\mu,\nu} (\tau) = i \frac{M_{\mu,\nu}}{\Delta_{\mu,\nu}} \left(\nep^{-i\tau\epsilon_\nu/\hbar} - \nep^{-i\tau\epsilon_\mu/\hbar} \right) |\phi_\mu\rangle \langle \phi_\nu | \ .
\end{equation}
Since we are interested in computing the projection $\langle \phi_\nu| u_\nu \rangle $, only the terms proportional to $|\phi_\nu \rangle $ are needed. 
Clearly Eq.~\eqref{eq:corr1} gives no contribution --- all terms are orthogonal to $|\phi_\nu \rangle $--- but it can be used to obtain the next order by imposing the normalization condition $\langle u_\nu | u_\nu \rangle =1$ 
\begin{equation}
\begin{split}
|u_\nu^{(2)}\rangle = &- \frac{\hbar^2}{2\tau^2} |\phi_\nu \rangle \sum_{\mu \neq \nu} 
\vert \frac{M_{\mu,\nu}}{\Delta_{\mu,\nu}} \vert ^2 \\
 &\hspace{0mm} + \mathrm{terms\ orthogonal\ to\ } |\phi_{\nu} \rangle \ . 
\end{split}
\end{equation}
Hence the occupation at finite frequency of the targeted Floquet mode reads
\begin{eqnarray}
n_{\nu} &=&\left| 1-\frac{\hbar^2}{2\tau^2} \sum_{\mu \neq \nu} 
\vert \frac{M_{\mu,\nu}}{\Delta_{\mu,\nu}} \vert ^2 \right|^2 + o(1/\tau^2) \nonumber \\
&=& 1-\frac{\hbar^2}{\tau^2} \sum_{\mu \neq \nu} 
\vert \frac{M_{\mu,\nu}}{\Delta_{\mu,\nu}} \vert ^2 + o(1/\tau^2) \;.
\end{eqnarray}
Therefore, if the matrix elements $M_{\mu,\nu}$ are not all equal to zero, we expect to see power law correction to the occupation number of Floquet modes,
when the system is prepared in the $\nu-$th state $|\phi_\nu(0)\rangle $ of $\hat{H}(t=0)$.

\section{Dependence on $k_x$ of Floquet quasi-energies and occupations} \label{app:Floquet}
Here we discuss the dependency of Floquet modes and quasi-energies form $k_x$ and how the system can be effectevely described in only 1+1 dimensions (space + time).
The starting point is the block diagonal Hamiltonian in momentum space, which reads 
\begin{equation}\label{eq:appHk}
\begin{split}
 \hat{H}_{\mathbf{k}}(t)=J_0 \sum_{b=0}^{q-1} 
\bigg\{ 2 \cos\left( ak_y + {\scriptstyle{\frac{2\pi p}{q}}} b \right) \opcdag{\k,b} \opc{\k,b}+ \\
\Big[  \nep^{-ia(k_x+\kappa_x(t))} \opcdag{\k,b+1} \opc{\k,b} + {\mathrm H.c.}  \Big] \bigg\} \ ,
\end{split}
\end{equation}
where $\opc{\k,q}=\opc{\k,0}$ and $a\kappa_x(t)=\omega (t-t_{\on}) $ when the force field $F=\frac{\hbar \omega}{a}$ is stationary. 
$t_{\on}$ is the initial time for which the system is prepared with a non periodic driving.
Notice that $\hat{H}_{\k}(t)$ depends on $k_x$ and time only through the phase $ak_x+\omega (t-t_{\on})$.
Hence we can define $t_x=t_{\on}-k_x/\omega$, so that the evolution operator over one period (the Floquet operator) for a given $t_{\on}$ can be written as
\begin{equation}
\hat{F}_{\k}(\tau)=\hat{U}_\k (\tau+t_{\on},t_{\on})=\hat{U}_{k_y}(\tau+t_x, t_x) \ ,
\end{equation}
Here a subscript $k_y$ indicates that the associated quantity is evaluated in $\k=(0,k_y)$.
By applying the composition property of evolution operator and exploiting  Floquet theorem in the form 
\[\hat{U}(t_{\on}+t+\tau,t_{\on})=\hat{U}(t+t_{\on},t_{\on})\hat{U}(\tau+t_{\on},t_{\on}) \ , \] one obtains
\begin{equation}
\hat{F}_{\k}(\tau)=\hat{U}_{k_y} (t_x,t_{\on})\hat{F}_{k_y}(\tau) \hat{U}^\dagger_{k_y} (t_x,t_{\on}) \ ,
\end{equation}
which can be written explicitely as 
\begin{equation}\label{eq:Floqkx}
\hat{F}_{\k}(\tau)=\sum_\nu \nep ^{-i \epsilon_{k_y,\nu} \tau/\hbar} |u_{k_y,\nu}(t_x)\rangle \langle u_{k_y,\nu}(t_x)| \ ,
\end{equation}
since the phase factors arising from the action of $\hat{U}_{k_y} (t_x,t_{\on})$ and $\hat{U}^\dagger_{k_y} (t_x,t_{\on})$ exactly cancel each other.
Therefore the Floquet modes shifted along $k_x$ are 
\begin{equation}\label{eq:ukx}
\begin{split}
|u_{\k,\nu}(t_{\on})\rangle &= |u_{k_y,\nu}(t_{\on}-\frac{ak_x}{\omega})\rangle \\
&=\nep^{i\epsilon_{k_y,\nu}(t_x-t_{\on})/\hbar} \hat{U}_{k_y}(t_x,t_{\on})|u_{k_y,\nu}(t_{\on})\rangle \ ,
\end{split}
\end{equation}
i.e. the periodic part of the $\nu$-th Floquet state in $\k=(0,k_x)$ evolved for a time $t_x-t_{\on} < \tau$.
Thus the Floquet operator at any point in the $k-$ space with $k_x\neq 0$ can be obtained by a {\em unitary} transformation applied on $	\hat{F}_{k_y}$.
The most important implication is that the quasi-energies $\epsilon_{\k,\nu}=\epsilon_{k_y,\nu}$ are independent from $k_x$.
The Floquet modes instead still depend on $k_x$, because of Eq.~\eqref{eq:ukx}.
Hence when computing the infinite time average pumped charge
\begin{equation}
Q_{\rm d} = \frac{\tau}{\hbar} \sum_{\nu} \int_{\rm BZ} \frac{\ud^2 \k}{(2\pi)^2} \; 
n_{\k,\nu} \frac{\partial \epsilon_{k_y,\nu}}{\partial k_y} \ ,
\end{equation}
the only remaining dependence on $k_x$ is in the occupation number $n_{\k,\nu}=|\langle \psi_\k| u_{k_y,\nu}(t_x) \rangle |^2$.

Another interesting property of the Hamiltonian as written in Eq.~\eqref{eq:appHk}, is that the spectrum is
invariant for a discrete shift of the momentum in the $\hat{{\bf y}})$ direction $k_y\rightarrow k_y + \frac{2\pi p}{qa}$.
Indeed this transformation is equivalent to a shift of $a$ in real space of the magnetic unit cell, leading to a simple relabelling of the internal index $b \rightarrow b +1$. 
For the case under investigation $(p=1$,  $q=3)$, this property is clearly shown in Fig.~\ref{fig:bands-chern}, where the invariance for $k_y \rightarrow k_y + \frac{2\pi}{3a}$ is evident.
This symmetry in the Hamiltonian is inherited also by the quasi-energy spectrum, which is also repeated three times inside the Brillouine zone.
This symmetry is nothing else than gauge invariance: the spectrum must depend on the same way by $k_x$ and $k_y$,
because the braking of translational invariance along the $\hat{\rm {\bf x}}$ direction is only due to the gauge choice, which can not influence any observable.
If we had chosen ${\bf A}= -By \hat{\x} $, the magnetic unit cell would have consisted of three sites along the $\hat{\y}$ direction and thus the first Brillouine zone would have been $[0,\frac{2\pi}{a})\times [0, \frac{2\pi}{3a})$, 
leading to a periodicity of $ \frac{2\pi}{3a}$ in $k_y$.


\end{document}